\documentclass{article}
\usepackage{graphicx}
\usepackage{enumerate}
\usepackage[square]{natbib}
\usepackage{authblk} 
\date{\today}

\begin{document}

\title{Ockham\textsc{\char13}s razor and the interpretations of quantum mechanics}
\author[1]{Gerd Ch. Krizek}
\affil{University of Vienna - Faculty of physics - Quantum particle group, Boltzmanngasse 5, 1090 Vienna, Austria.}
\affil{UAS Technikum Wien - Department of Applied Mathematics and Sciences, H\"ochst\"adtplatz 6, 1200 Vienna, Austria.}


\maketitle

\begin{abstract}
\noindent 
Ockham\textsc{\char13}s razor is a heuristic concept applied in philosophy of science to decide between two or more feasible physical theories. Ockham\textsc{\char13}s razor operates by deciding in favour of the theory with least assumptions and concepts; roughly speaking, the less complex theory. Could Ockham\textsc{\char13}s razor not easily treat the different interpretations as theories and decide in favour of the one with fewest assumptions? We provide an answer to this question by means of examples of applications in literature and the discussion of its historical origin. We review the historical context of Ockham\textsc{\char13}s razor and its connection to medieval philosophical struggles, discuss the essence of its parsimonious core and put it in relation with modern struggles in the context of interpretational issues in Quantum Mechanics. Due to the lack of experimental evidences in string theory, a new field of modern heuristics arose in the last years. We will discuss these heuristics in the context of Ockham-related heuristic methods and analyse the connection of these heuristics with the interpretations of Quantum Mechanics. 

\end{abstract}

\newpage
\section{Introduction}
\parindent 0cm       
The interpretations of Quantum Mechanics are up to now matter of philosophical-physical debates. One philosophical concept that is often stressed during debates about interpretational issues in Quantum Mechanics is Ockham\textsc{\char13}s \footnote{In this work, we will use the wording ``Ockham'' instead of ``Occam''. In the literature, both can be found.} razor, or the principle of parsimony. \newline

\cite{gu2011occam} give a good example how Ockham\textsc{\char13}s razor is usually perceived in science:

\begin{quote}
\textit{``Occam\textsc{\char13}s razor, the principle that plurality is not to be posited without necessity, is an important heuristic that guides the development of theoretical models in quantitative science."} 
\end{quote}  

To get a deeper understanding about the significance of Ockham\textsc{\char13}s razor, an overview about the historical context and environment is needed to understand the origins of his razor principle. Ockham\textsc{\char13}s comprehensive work covers logic, theology, politics, epistemology and philosophy, although today he is best known for his parsimony principle. With his philosophical position, nowadays called nominalism\footnote{We use the term Nominalism for Ockham\textsc{\char13}s position on the problem of universals. Nominalism is usually a much broader concept and contains several discriminable ideas\cite[]{beckmann1998ockham}; \cite[]{goodman1947steps}. }, he encountered the problem of universals, which dominated the philosophy of the Middle Ages. The problem of universals is the question whether a general property or quality, like the colour ``red'' or the term ``apple,'' has an existence on it´s own or if it is only a name for a certain property of an individual object \cite[]{rodriguez2000problem}. We will discuss the problem of universals in more detail at a later stage of this work, but one can already safely say that it still has relevance in modern philosophy; for example, in the philosophy of mathematics, where we operate with numbers as philosophical entities. 
\newline

%
%

In Quantum Mechanics, Ockham\textsc{\char13}s razor has been used in literature in the interpretational debates as reason to favour specific interpretations. We will give an insight into the historical and philosophical environment where Ockham\textsc{\char13}s razor had its origins, and discuss how the principle usually is used in reasoning in the quantum-mechanical context to discuss its applicability in Quantum Mechanics.

First of all, we would like to clarify a still widely spread mistake. The usually cited phrases for Ockham\textsc{\char13}s razor concept are not by William of Ockham, they are artifacts of translations, summarizations and simplifications done by successors. A detailed analysis of this myth around Ockham\textsc{\char13}s razor has been given by \cite{thorburn1918myth}. He gives a comprehensive collection of all original quotations concerning Ockham\textsc{\char13}s parsimony principle, for which we will provide a translation and detailed discussion in section 2.3.

\newpage

\section{Ockhams philosophy in historical context}
In this section, we provide an overview of the historical context and environment to understand the origins of Ockham\textsc{\char13}s razor principle. William of Ockham was a Franciscan friar that lived between 1287 and 1347. He received a theological education in the Franciscan order at the London convent, and later in Oxford, but he never completely finished with a master of theology, though he fulfilled all requirements. It is believed that this was due to his controversial views, which antagonized some of his fellows and superiors, and to some conflicts between the Franciscan order and Pope John XXII. 

Following \cite{beckmann1995wilhelm}, Ockham\textsc{\char13}s philosophy is pervaded by three major principles: The omnipotence principle, the contradiction principle, and the parsimony principle. All three of them have strong interrelations and a deep theological connection. This theological context plays no role for our application of Ockham\textsc{\char13}s razor onto physics, but is important to understand Ockham\textsc{\char13}s epistemological ideas in the context they were created. 

It is important to note that these principles have never been set and classified by Ockham himself, and this classification is not used in literature universally. Nevertheless, these principles provide a good overview of the ideas of Ockham, which is the reason for us to make use of them here.\footnote{By analysing Ockham\textsc{\char13}s original writings, it is clear that no such structure of principles is defined there. \cite[]{beckmann1995wilhelm} emphasizes that as well}

\subsection{The omnipotence principle}

The omnipotence principle describes the omnipotence of the theological entity god and his unrestricted ability to act as he wishes to do. In Ockham\textsc{\char13}s own words \cite[OT VII, 45]{william2012william}:

\begin{quote}
\textit{``Deus nullus est debitor.'' or ``God is no one\textsc{\char13}s debtor.''\footnote{Translation by Gerd Ch. Krizek} }

\end{quote} 

Following \cite{beckmann1995wilhelm} Ockham\textsc{\char13}s god is nearly unrestricted in his actions to create the world according to his will and free to change it as he wishes. Although the theological entity god cannot act completely unrestricted, there is one restriction: god is not able to act contradictory, and nothing in his creation is allowed to be contradictory. 
With this idea Ockham encountered the philosophy of necessitarianism, which had its origins in the work of Aristotle and Platon, where all things exist due to necessity and therefore our world as it exists is the only possible one. In a theological context, Ockham\textsc{\char13}s position was radical and new, and was one of several reasons for Ockham\textsc{\char13}s conflict with censorship and the papal court \cite[]{beckmann1998ockham}; \cite[]{klocker1996william}.   

As mentioned before, the restriction of the entity god is contradiction: this is settled in the second principle. For further details on this topic, refer to \cite[]{schrocker2003verhaltnis}.

\newpage

%
%
%
%
%
%

\subsection{The contradiction principle}
\label{sec:contradiction}

Ockham\textsc{\char13}s contradiction principle is a continuation of the aristotelian theory of thought. Aristotle formulated this first principle in the following way \cite[]{ross1928works}:
\begin{quote}
\textit{``It is, that the same attribute cannot at the same time belong and not belong to the same subject and in the same respect.''}
\end{quote}
 
In Ockham\textsc{\char13}s epistemology, the idea is continued in the sense that contradictions cannot even be purpose of knowledge and perception. Perceptibility and contradictoriness exclude each other in principle, according to Ockham. The striking point is not that this idea is new, but that this is Ockham\textsc{\char13}s starting point to aim for nominalism; this is his way to handle the problem of universals.

As outlined before, the problem of universals is the question if wether general terms like the colour ``red'' have an ontological existence on their own. If this would be affirmed, ``redness'' would be a qualia on its own. 
The problem of universals got discussed over the centuries until even now. Famous participants were Kant, Peirce, Russel, Whitehead and Wittgenstein \cite[]{kant1889immanuel};\cite[]{peirce1974collected};\cite[]{whitehead1912principia};\cite[]{wittgenstein1994tractatus}. 
\newline
Ockham handles the problem of universals in his comments on the ``Four books of Sentences'' by Peter Lombard, which was a standard exercise for every medieval theologian. The scholastic tradition dominated the medieval theology. It was based on the dialectical method, where a question is formulated, arguments against and in favour of a statement were discussed and replies to the objections were given. An analysis of this scholastic method, as an example by Ockham himself, can be found in Appendix A.

\bigskip

With his nominalist position Ockham argues that the objects of our perception in the world are individuals; they are unique and it is not necessary to use the concept of universals to describe these objects. Moreover, the universals are not entities that are necessarily needed; they are connected to the construction of language and the way how the human mind perceives and processes the world. The antithetic concept to nominalism is realism. It is important to distinguish this definition of realism, which only refers to the dispute on universals, from the realism concepts in philosophy of science such as the naive realism or the scientific realism. The realists position concerning the problem of universals assigns an objective existence to universals and goes back to the ideas of Platon. Therefore, the realist position is often called Platonism or Platonic realism, even though these concepts have several differences. Regarding the problem of universals, their position is identical; thus, we will stick to this naming and subsequently refer to this position as Platonic realism.

Now we will show how the contradiction principle plays an important role in Ockham\textsc{\char13}s reasoning against Platonic realism. \newline

\newpage

Following \cite[p.105]{schonberger1990realitat} Ockham\textsc{\char13}s way of arguing is like this: 

\begin{quote}
\textit{``2 contradictory statements, if they should both be true at the same time, can only refer to two different things. And vice versa: Wherever something distinguishable occurs, contradiction can be claimed. Therefore a designated contradiction, if its contradictory contents can refer to distinguishable things, is a ground-breaking argument for Distinguishability or Difference.''}
\end{quote}

Or with Ockham\textsc{\char13}s own words \cite[OTh II, Liber I, D.2 Q.6, page 174]{william2012william}:

\begin{quote}
\textit{``contradictio est via potissima ad probandum distinctionem rerum.''  
\newline
or
\newline ``Contradiction is the most important way to prove difference.''\footnote{Translation by Gerd Ch. Krizek}}
\end{quote}

The contradiction principle should be understood as an attack on the concept of universals and intercessor of nominalism. Let us discuss how this is done with the help of a Gedankenexperiment.
  
\begin{figure}[h]

\begin{center}
\includegraphics[width=0.5\textwidth]{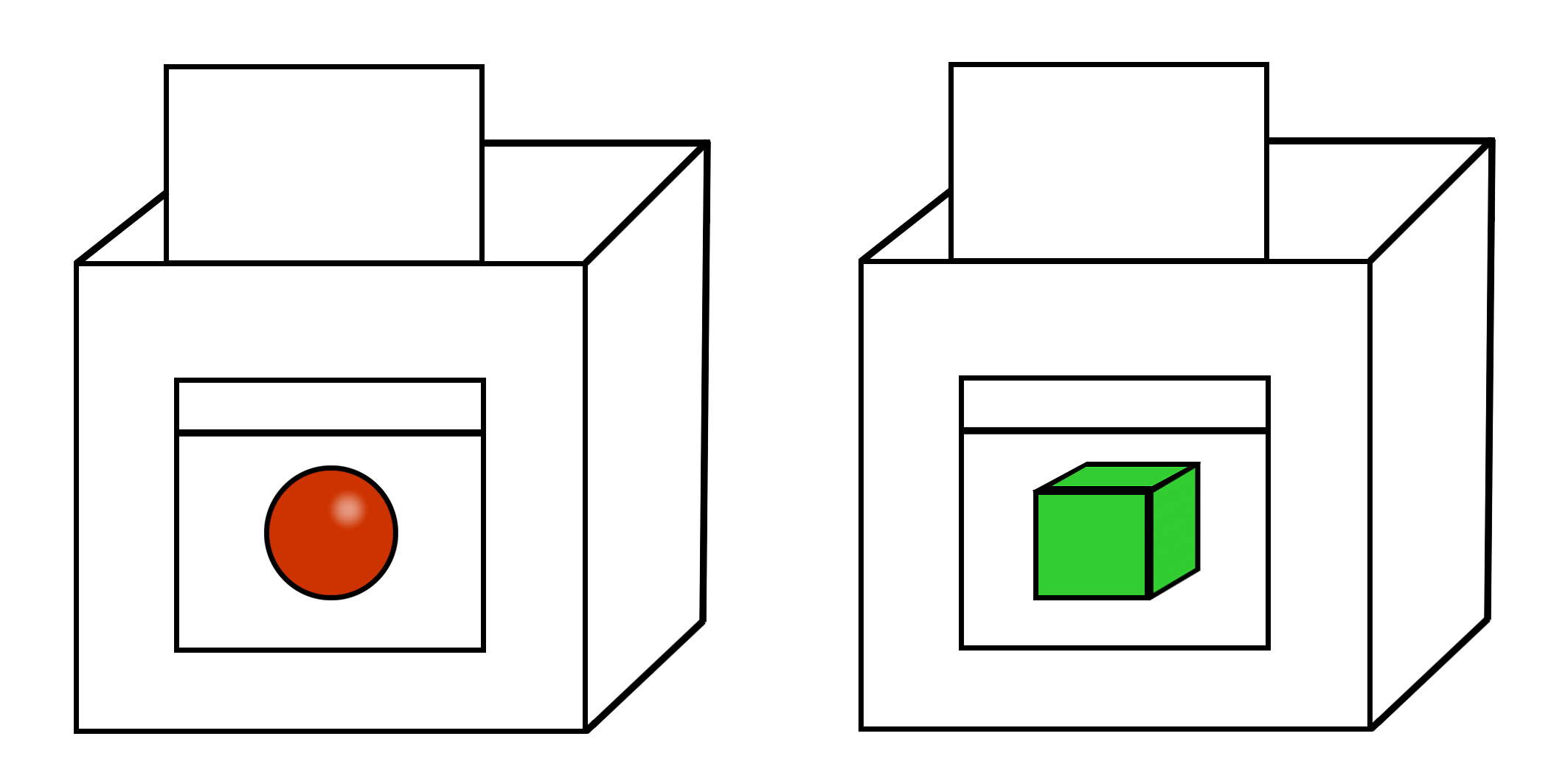}
\caption{The objects in the boxes} 
 \label{Fig1}
\end{center}

\end{figure}

Let us assume a box where a person examines the content of the box two times. In between the process of looking for the properties of the content, it should be possible that someone secretly replaces the objects in the box with different ones. The question is if whether the objects in the box at the two times are identical. The objects we assume in our Gedankenexperiment have properties A and B; for convenient illustration, let it be colour and shape. Furthermore, we assume both properties A and B are dichotomic, so for colour there are only to possible states red or green; for shape let it be round or angled. 

\newpage

The following Illustration shows the properties (qualia) of the assumed objects.

\begin{figure}[h]

\begin{center}
\includegraphics[width=0.5\textwidth]{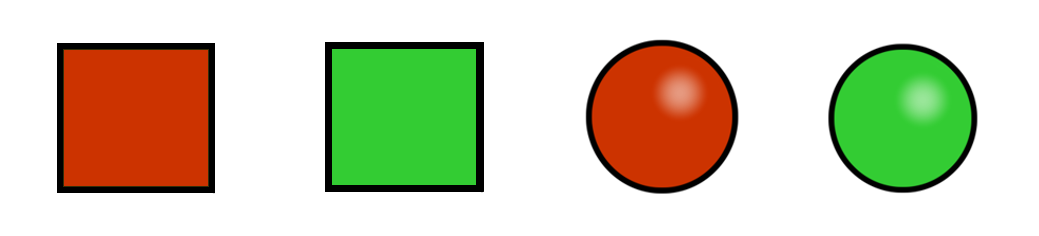}
\label{Box}
\caption{The qualia of the objects}
 
\end{center}

\end{figure}

If the person controls the boxes twice, as defined, out of the 16 possible combinations, the following cases might occur:

\begin{figure}[h]

\begin{center}
\includegraphics[width=1\textwidth]{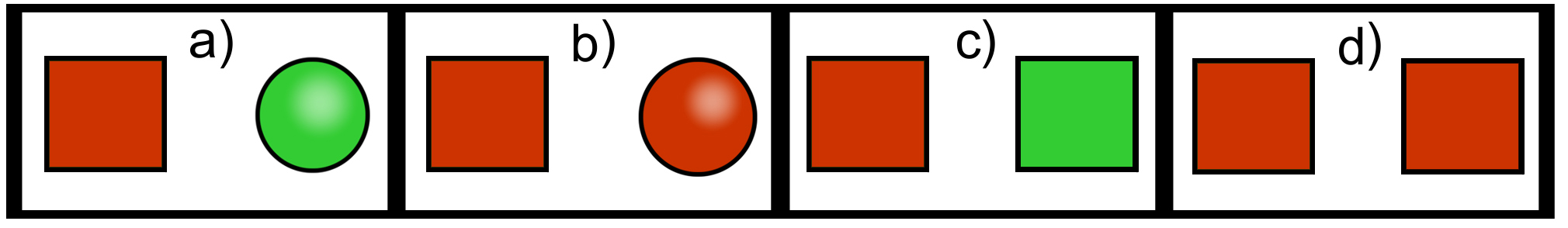}
\caption{4 of 16 possible combinations}
 \label{Fig2}
\end{center}

\end{figure}

\begin{enumerate}[(a)]

\item
\begin{eqnarray}
\centering \nonumber
colour_{1}=red && colour_{2}=green \nonumber \\ \nonumber
shape_{1}=angled && shape_{2}=round 
\end{eqnarray}

\begin{eqnarray}
colour_{1}\ne colour_{2} \nonumber \\
shape_{1}\ne shape_{2} \nonumber \\
\hookrightarrow DIFFERENCE \nonumber
\end{eqnarray}

\item
\begin{eqnarray}
\centering \nonumber
colour_{1}=red && colour_{2}=red \nonumber \\ \nonumber
shape_{1}=angled && shape_{2}=round 
\end{eqnarray}

\begin{eqnarray}
shape_{1}\ne shape_{2} \nonumber \\
\hookrightarrow DIFFERENCE \nonumber
\end{eqnarray}

\item
\begin{eqnarray}
\centering \nonumber
colour_{1}=red && colour_{2}=green \nonumber \\ \nonumber
shape_{1}=angled && shape_{2}=angled 
\end{eqnarray}

\begin{eqnarray}
colour_{1}\ne colour_{2} \nonumber \\
\hookrightarrow DIFFERENCE \nonumber
\end{eqnarray}

\item
\begin{eqnarray}
\centering \nonumber
colour_{1}=red && colour_{2}=red \nonumber \\ \nonumber
shape_{1}=angled && shape_{2}=angled 
\end{eqnarray}

\begin{eqnarray}
No \quad contradiction \nonumber \\
\hookrightarrow IDENTITY  \nonumber
\end{eqnarray}

\end{enumerate}

By the fact that there is a contradiction in the statements about the properties of the objects in the box at the two times, difference for the objects can be concluded. The conclusion is: Because of the contradiction, difference can be supposed. On the other hand, by perceiving no contradiction in the properties, it is reasonable to conclude identity for the objects.

In this last example, we avoided making use of universals; now let\textsc{\char13}s introduce the concept of universals and add them to our Gedankenexperiment. Additional to our dichotomic properties A and B, we assume that the universal ``toybrick'' should be applicable to all of our objects. It is important to see that univerals are not properties in general and therefore not per se perceptible to our observer.    

\begin{enumerate}[(a)]

\item
\begin{eqnarray}
\centering \nonumber
colour_{1}=red && colour_{2}=green \nonumber \\ \nonumber
shape_{1}=angled && shape_{2}=round \\
universal_{1} = toybrick && universal_{2} = toybrick \nonumber
\end{eqnarray}

\begin{eqnarray}
colour_{1}\ne colour_{2} \nonumber \\
shape_{1}\ne shape_{2} \nonumber \\
\hookrightarrow DIFFERENCE \nonumber
\end{eqnarray}

\item
\begin{eqnarray}
\centering \nonumber
colour_{1}=red && colour_{2}=red \nonumber \\ \nonumber
shape_{1}=angled && shape_{2}=round \\
universal_{1} = toybrick && universal_{2} = toybrick \nonumber
\end{eqnarray}

\begin{eqnarray}
shape_{1}\ne shape_{2} \nonumber \\
\hookrightarrow DIFFERENCE \nonumber
\end{eqnarray}

\item
\begin{eqnarray}
\centering \nonumber
colour_{1}=red && colour_{2}=green \nonumber \\ \nonumber
shape_{1}=angled && shape_{2}=angled \\
universal_{1} = toybrick && universal_{2} = toybrick \nonumber
\end{eqnarray}

\begin{eqnarray}
colour_{1}\ne colour_{2} \nonumber \\
\hookrightarrow DIFFERENCE \nonumber
\end{eqnarray}

\item
\begin{eqnarray}
\centering \nonumber
colour_{1}=red && colour_{2}=red \nonumber \\ \nonumber
shape_{1}=angled && shape_{2}=angled \\
universal_{1} = toybrick && universal_{2} = toybrick \nonumber
\end{eqnarray}

\begin{eqnarray}
No \quad contradiction \nonumber \\
\hookrightarrow IDENTITY  \nonumber
\end{eqnarray}

\end{enumerate}

As we see, adding the universal ``toybrick'' makes no difference to our conclusions on difference and identity. The reason is that all objects were attributed with the same universal. Let\textsc{\char13}s assume that two toybrick manufacturers produce toybricks with our binary dichotomic properties A and B, colour (in red and green) and shape. Let them be LEGO\copyright{} and K´NEX\copyright{}. Now let us assume that the vicious person that wants to puzzle our experimenter, who is examining the properties of the objects in the boxes, always replaces a LEGO\copyright{} toybrick at the first check with a K´NEX\copyright{} toybrick for the second examination.

\begin{enumerate}[(a)]

\item
\begin{eqnarray}
\centering \nonumber
colour_{1}=red && colour_{2}=green \nonumber \\ \nonumber
shape_{1}=angled && shape_{2}=round \\
universal_{1} = LEGO\copyright{} && universal_{2} = K´NEX\copyright{} \nonumber
\end{eqnarray}

\begin{eqnarray}
colour_{1}\ne colour_{2} \nonumber \\
shape_{1}\ne shape_{2} \nonumber \\
\hookrightarrow DIFFERENCE \nonumber
\end{eqnarray}

\item
\begin{eqnarray}
\centering \nonumber
colour_{1}=red && colour_{2}=red \nonumber \\ \nonumber
shape_{1}=angled && shape_{2}=round \\
universal_{1} = LEGO\copyright{} && universal_{2} = K´NEX\copyright{} \nonumber
\end{eqnarray}

\begin{eqnarray}
shape_{1}\ne shape_{2} \nonumber \\
\hookrightarrow DIFFERENCE \nonumber
\end{eqnarray}

\item
\begin{eqnarray}
\centering \nonumber
colour_{1}=red && colour_{2}=green \nonumber \\ \nonumber
shape_{1}=angled && shape_{2}=angled \\
universal_{1} = LEGO\copyright{} && universal_{2} = K´NEX\copyright{} \nonumber
\end{eqnarray}

\begin{eqnarray}
colour_{1}\ne colour_{2} \nonumber \\
\hookrightarrow DIFFERENCE \nonumber
\end{eqnarray}

\item
\begin{eqnarray}
\centering \nonumber
colour_{1}=red && colour_{2}=red \nonumber \\ \nonumber
shape_{1}=angled && shape_{2}=angled \\
universal_{1} = LEGO\copyright{} && universal_{2} = K´NEX\copyright{} \nonumber
\end{eqnarray}

\begin{eqnarray}
No \quad contradiction \nonumber \\
\hookrightarrow IDENTITY  \nonumber
\end{eqnarray}

\end{enumerate}

\newpage 

We see, in fact, that in fact the universal assigned to an object with identical properties might be different, but there is no way to distinguish them. Therefore in Ockham\textsc{\char13}s view it makes no sense to speak about the reality of those universals; they do not seem connected to the real world, if they do not reside in distinguishable properties of the objects. Furthermore, it would be possible to multiply them easily. If the toybrick belongs to Sarah, you could assign it a new universal, ``Sarah\textsc{\char13}s toybrick,'' and so on. This numberless amount of universals that would then reside in reality, in the view of Platonic Realism, was unneccessary to Ockham. It is not useful to introduce more entities than necessary. 
\newline
\newline
This is the essence of Ockham\textsc{\char13}s Parsimony principle, and one recognizes the strong connection to the Contradiction principle. In Ockhams own words

\cite[OP I,SL I,12]{william2012william} or here \cite[SL I,12 page 58]{Ockh1}: 
\begin{quote}
\textit{``Frustra fit per plura quod fieri potest per pauciora.''}
\newline
\textit{``For nothing is done with a multiplicity, what can be done with less.''\footnote{Translation by Gerd Ch. Krizek}}
\end{quote}

 We will discuss the Parsimony principle in more detail in section 2.3. 
\newline


\newpage

\subsection{The parsimony principle}

The parsimony principle is the heuristic method that is meant when people speak of Ockham\textsc{\char13}s razor. There are several formulations in the literature which are not originally by Ockham himself. Furthermore, it is not the case that there is a specific writing of Ockham where he states this principle as fundamental and assigns deeper relevance to it (although it is anchored in his whole thinking and pervades his philosophical and theological works). One original formulation can be found here \cite[OP I,SL I,12]{william2012william} or here \cite[SL I,12 p.58]{Ockh1}: 
\begin{quote}
\textit{``Frustra fit per plura quod fieri potest per pauciora.''}
\end{quote}

We may translate this sentence as

\begin{quote}
\textit{``For nothing is done with a multiplicity, what can be done with less.''\footnote{Translation by Gerd Ch. Krizek}}
\newline
\newline
or in more modern words
\newline
\newline
\textit{``It is pointless to do with more what can be done with fewer.''\footnote{Translation by Gerd Ch. Krizek}}
\end{quote}


Several other formulations used in modern literature are not by Ockham himself. \cite{thorburn1918myth} clarified the question around the formulation and gives a comprehensive overview about formulations of the parsimony principle, but only in Latin. To illustrate the idea in its historical context, it makes sense to give translations\footnote{Translation Josef Reiter and Gerd Ch. Krizek}:

\begin{enumerate}[(a)]
         \item 
         \begin{quote}
         \textit{``A multiplicity may never be stated without necessity.''}
         \end{quote}
         
         \item 
         \begin{quote}
         \textit{``...multiplicity may not be stated, but for it is necessary .''}
         \end{quote}
         
         \item          
         \begin{quote}
         \textit{``This opinion states a multiplicity without necessity, which is against the doctrine of philosophers.''}
         \end{quote}
         
         \item 
         \begin{quote}
         \textit{``As somebody who follows the common sense does not state multiplicity, but for that what is included by common sense, somebody that follows Believe may not state more than the truth of Believe requires.''}
         \end{quote}
         
         \item       
         \begin{quote}
         \textit{``The use of multiplicity must obviously always be necessary.''}
         \end{quote}
         
         \item                  
         \begin{quote}
         \textit{``For nothing is done with a multiplicity, what can be done with less.''}
         \end{quote}
         
         \item 
         \begin{quote}
         \textit{``Stringency should be posed, where multiplicity is not necessary.''}
         \end{quote}
         
      \end{enumerate}

Out of these passages and the context of the other principles we can conclude: 

\begin{itemize}
      \item As shown in Section \ref{sec:contradiction}, the contradiction principle plays an important role in Ockham\textsc{\char13}s arguments in favour of his nominalism against the existence of the universals, which to him are numberless in principle. They form the multiplicity he wanted to get rid of and where he applies the razor principle. 
      
      \item There is no statement against the existence of entities in general, nor does the parsimony principle represent the view that entities have to fulfill some extremum principle.
      
      \item The razor is no deep mystery to Ockham; it is just made of common sense. It is a very hands-on principle, a principle of economy of thought and resources.  
      
      \item It is a principle of adequacy. Parsimony means not to reduce to a minimum at all costs; it follows the “truth”, whatever that may be precisely in context of physics\footnote{In physics, the confirmation of the applicability of a certain model or theory is given by experiment. Nature only gives an answer to the question on adequacy of a theory. In context with experimentally not testable theories, some authors would see that statement as too strict. Examples on this point of view will be given in chapter \ref{sec:heuristics}}. Parsimony means to use as much as necessary and as less as possible.

\end{itemize}      

Out of this, we can conclude that Ockham\textsc{\char13}s razor has no meaning as a strict ontological principle; it is not meant to be applied to ontological entities. It is an epistemological statement to minimize the number of symbols used to describe a certain sensational experience. 

It is again Wittgenstein who sums up the razor principle \cite[ 5.47321 and 3.328]{wittgenstein1994tractatus} in a comprehensible way:

\begin{quote}
\textit{``Occam’s maxim is, of course, not an arbitrary rule, nor one that is justified by its success in practice: its point is that unnecessary units in a sign-language mean nothing.''}
\end{quote}

\begin{quote}
\textit{``If a sign is useless, it is meaningless. That is the point of Occam’s maxim.''}
\end{quote}

To Ockham, the universals represent such useless symbols that do not reside in reality.

The notion of simplicity is strongly connected to Ockham\textsc{\char13}s razor. It is often referred to as a principle of simplicity. That is somehow misleading and should be discussed in more detail. Simplicity is a very subjective criterion unless it is somehow quantified by a measure. Often Ockham\textsc{\char13}s razor is understood as the concept that only simple explanations are the correct ones; in the context of a theory, the simplest theory must be right. If that would be the case, in a naive way it would be pretty easy to do science, but this is obviously not the case. The simplicity goes only as far as possible. Adequacy and consensus with the experiment are favored over simplicity. Physics does not describe the world itself, it describes mental representations of experiences we conduct in the real world. Therefore, the simplicity, if applicable, does not belong to the world, but to the mental representation of the experiences. It is a comprehensible concept that human minds select the simplest way to describe an experience by mental representations. This concept is economical and parsimonious concerning the limited resources of the human mind\footnote{For further details refer to \cite[]{tversky1973availability}}, but it does not necessarily tells us more about the things in reality. The success of science and its miracle, if you would like to use the notion of the miracle argument, is that our mental representations deliver models with predictive power.  \newline

Wittgenstein emphasizes the subjective element of the parsimonious method \cite[6.363 and 6.3631]{wittgenstein1994tractatus}:

\begin{quote}
\textit{``The procedure of induction consists in accepting as true the simplest law that can be reconciled with our experiences. \newline
This procedure, however, has no logical justification but only a psychological one.''}
\end{quote}

Others emphasize the connection between simplicity and truth. \cite{kelly2010simplicity} formalises the concept of Ockham\textsc{\char13}s razor for truth finding and sets up an Ockham efficiency argument. A detailed analysis of this ideas would go beyond the scope of this work, but it can be seen ad hoc that the notion of truth is something difficult in context with physics and science in general, and probably somehow confined to some technical applications like machine learning, where it is clear what is defined by the concept of truth. 

We will be concerned with the notion of simplicity again in the context of the interpretations of Quantum Mechanics, but finally we will present a biological application of Ockham\textsc{\char13}s razor and the notion of simplicity. \newline

\cite{westerhoff2009systems} argues in his justification for Systems Biology that 
\begin{quote}
\textit{``Two paradigms, i.e. Ockham\textsc{\char13}s razor and the prevalence of minimum energy solutions are pertinent to much of physics ''} 
\end{quote} 
    
and referring to living organisms:

\begin{quote}
\textit{``For researching living organisms, Occams razor is not an appropriate paradigm.''} 
\end{quote} 

... due to the high complexity and the high amount of constituents involved. He somehow argues that you can speak about simplicity in the scale of ...

\begin{quote}
\textit{``two or three components, and if that does not work perhaps 6. Three hundred and seventy-five is certainly not in the realm of simplicity.''} 
\end{quote} 

The application of Ockham\textsc{\char13}s razor in the regime of things in reality\footnote{Like a biological system} is not the intended application by Ockham. Ockham\textsc{\char13}s arguments are applicable for mental representations that should fulfill the parsimony, not the things itself. The application of Ockham\textsc{\char13}s razor as a concept of parsimony or simplicity in the things itself goes far beyond the ideas of Ockham, therefore, it would be better to name efficiency concepts in different regimes than that intended by Ockham, differently than ``Ockham\textsc{\char13}s razor''.  
\newline\newline
It would be an over simplification to reduce Ockham\textsc{\char13}s thinking to a few principles and fairly simple conclusions; it would not meet Ockham\textsc{\char13}s philosophy in all its facets and subtleties, but for our purpose of a modern context it seems sufficient and appropriate to demonstrate his ideas in context of those principles. Further subtleties and the connection to the theological aspects\footnote
{
The contradiction principle has one more aspect worth mentioning, which is negligible for the physical context of Ockham\textsc{\char13}s razor, but still interesting for the historical embedding of Ockham\textsc{\char13}s philosophy and theology. The contradiction principle is in connection with the omnipotence principle in the context that it even is impossible for the entity god to act in contradictory ways.\newline
A modern account of this idea has been given by \cite[3.031]{wittgenstein1994tractatus}:
\begin{quote}
\textit{``It used to be said that God could create anything except what would be contrary to the laws of logic. The truth is that we could not say what an ``illogical'' world would look like.''}
\end{quote}
This principle was stressed mainly during disputes about Trinity, because of its inherent logical contradiction \cite[]{beckmann1998ockham}.
} 
are discussed extensively in \cite[]{schonberger1990realitat}. For logical aspects of the contradiction principle, refer to \cite[]{schick2010contradictio}.

\newpage

\section{Ockham and other heuristics}
\label{sec:heuristics}

Before discussing the role of Ockham\textsc{\char13}s razor in the interpretations of Quantum Mechanics, we will give an overview over other heuristics connected to Ockham\textsc{\char13}s razor in recent physical debates. \newline
 
No claim is made on completeness of the selection of heuristics presented here; they have been chosen due to relevance for our discussion.  

\subsection{Machs economy of thought}

Ernst Mach pleads for an idea of economy of thought, which emerged from his works in philosophy of science. He takes the phenomenalistic point of view that all our scientific reasoning is based on economy of thought, starting with language, perception, and mathematics pervading all fields of scientific work. By perceiving sensations of things in the world, classifying them into boxes\footnote{The boxes mean the categories in which the sensations of things are sorted} and constructing laws out of their behavior, one does not gain knowledge about the things itself, one merely connects the mental representations of the things with laws \cite[]{mach1898popular}:

\begin{quote}
\textit{``It thus comes to pass that we form the notion of a substance distinct from its attributes, of a thing-in-itself, whilst our sensations are regarded merely as symbols or indications of the properties of this thing-in-itself. But it
would be much better to say that bodies or things are compendious mental symbols for groups of sensations, symbols that do not exist outside of thought. 
Thus, the merchant regards the labels of his boxes merely as indexes of their contents, and not the contrary. He invests their contents, not their labels, with real value.''}
\end{quote}

His economy of thought now relates to the mind-based processes where the laws interconnect the mental representations of the things, the mental symbols, in an economical way. So the economic character comes in by the perceiving subject and its limitations. It is an economy of practicality, even in mathematics \cite[]{mach1898popular}:

\begin{quote}
\textit{``The greatest perfection of mental economy is attained in that science which has reached the highest formal development, and which is widely employed in physical inquiry, namely, in mathematics. Strange as it may sound, the power of mathematics rests upon its evasion of all unnecessary thought and on its wonderful saving of mental operations. Even those arrangement-signs which we call numbers are a system of marvellous simplicity and economy. ...
No one will dispute me when I say that the most elementary as well as the highest mathematics are
economically-ordered experiences of counting, put in forms ready for use.''}
\end{quote}

and further:

\begin{quote}
\textit{``Physics is experience, arranged in economical order.''}
\end{quote}

It can be clearly seen that the economy Mach proposes is connected to the limited capacity or ability of the perceiving subject. It is therefore a practical economy justified by human beings as operators of scientific methods, not an economy settled in the things as they are. Economy is understood as consequence of the development of predecessors of science, like handcraft or engineering, which made it necessary to transfer knowledge and experience. SI-units are human-centric from their choice of scale; likewise, the economic nature of physics resides in the economic nature of human language and thought. \newline
%
%

For a more detailed analysis and different aspects of Machs economy of thought, refer to \cite[]{kallfelz1929okonomieprinzip}.

\subsection{The No Alternative Argument}
\label{NAA}


The No Alternative Argument says that despite intensive research, no alternative theories have been found to a specific theory or theory complex, this gives an argument in favour of the specific theory. It sounds simple, but has been put on a formal foundation in \cite{dawid2014no}.\newline

 The argument of \citeauthor{dawid2014no} refers to string theory to give justification in absence of empirical evidences. It is part of the idea of non-empirical evidences and is connected to the Meta Inductive Argument and the Unexpected Explanation Argument. 

\subsection{The Meta Inductive Argument}
\label{MIA}


The Meta Inductive Argument is connected to the No Alternative Argument and Scientific Underdetermination, where it is argued that the number of possible alternative theories to a specific theory is limited in principle (Scientific Underdetermination \cite[]{dawid2007scientific} and \cite[]{dawid2008scientific}) and the No Alternative Argument favors the specific one when there is a lack of alternative theories. \newline

The Meta Inductive Argument says that if a specific theory has something in common with theories that have been empirically confirmed, it is reasonable to favor this specific theory \cite[]{dawid2014no}:

\begin{quote}
\textit{``Now, assume that a novel theory H shows similarities to theories H1,
H2, etc., in the same scientific research program. The joint feature of these
theories may be a certain theoretical approach, a shared assumption, or any
other relevant characteristic. Let us assume that a substantial share of the
theories to which H is similar have been empirically confirmed. Assume
further that for those theories, we have empirically grounded posterior beliefs
about the number of alternatives. Then, it seems reasonable to use these
posteriors as priors for the number of alternatives to H. After all, H is quite
similar to H1, H2, etc. In statistics, such a way of grounding ``objective''
prior beliefs in past experience is referred to as the empirical Bayes method''}
\end{quote}

For \cite{dawid2014no} the Meta Inductive Argument is not working alone; it has to be seen together with the No Alternatives Argument and the Unexpected Explanation Argument.

\subsection{The Unexpected Explanation Argument}
\label{UEA}


The Unexpected Explanation Argument is not an empirical argument, nor should it be mixed up with a novel empirical confirmation that supports a certain theory.  \newline

The Unexpected Explanation Argument claims that a theory explains a different theoretical content in a completely unexpected way or regime where the theory has not been developed for \cite[]{dawid2015modelling}:

\begin{quote}
\textit{``Theory H was developed in order to solve a specific problem. Once H was developed, physicists found out that H also provides explanations with respect to a range of problems which to solve was not the initial aim of developing the theory.''}
\end{quote}

The idea of Non-Empirical Confirmations refers to the arguments presented in \ref{NAA}, \ref{MIA} and \ref{UEA} and uses them in combination to give justification to theories that bear an inherent lack of empirical data. The arguments have been designed for use in context with string theory, but \cite{dawid2014no} argue for application in archaeology and palaeontology as well.

\subsection{The Ontological Coherence Argument (OCA)}

We would like to propose a new heuristic that might have been used, already implicit in history of science, the Ontological Coherence Argument. 
\newline
\newline
We claim that a theory consists of its mathematical apparatus, its physical interpretation by labels and names, its interpretation by concepts and principles and an ontological embedding. The details of the definition of these four levels of a physical theory will be discussed in a subsequent paper.
\newtheorem{OCA}{Theorem}
\begin{OCA}
Under the assumption of a set of Theories $ H$, that cover different regimes of physics, and that $ H$ are Theories in agreement with experimental data in the respective regime, the OCA claims that those Theories $ OT \in H$ whose ontologies fit coherently together are preferable. 
\end{OCA}


\begin{figure}[h] 

\begin{center}
\includegraphics[width=0.5\textwidth]{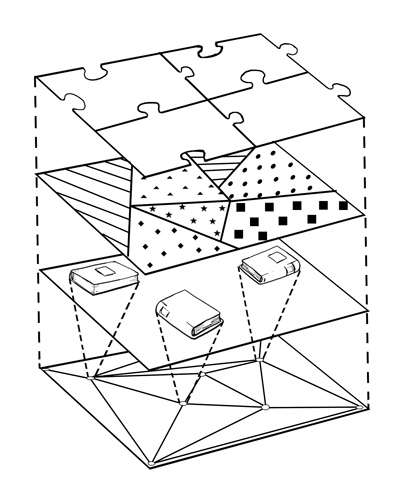}
\caption{Levels of interpretation}
\label{OCA}   
\end{center}

\end{figure}

The idea of OCA is not to rule out any discussed theory or interpretation within the framework of Quantum Mechanics, but to encourage to look out for connected theories outside of Quantum Mechanics to see if the puzzle of ontological entities gives rise to a bigger picture. A historical example where the application of the OCA can be seen applied is "The Great Cosmological Controversy" between the Geocentric, Heliocentric and Geoheliocentric model. Today it is widely unkown that there were striking empirical arguments in favour of the Geoheliocentric model by Tyho Brahe \cite[]{graney2010telescope}. The empirical arguments were telescopic measurements of star disks, which supported the Geoheliocentric model by the assumption that stars should all have the same order of size. It turned out later that the observed star disk was an optical artefact, an airy disk caused by diffraction. 
\newline
The picture of a new cosmology which connected to Isaac Newton\textsc{\char13}s Principia Mathematica was ontologically striking; the Heliocentric view connected seamless to the nature of gravity and the laws of motion. Not only was it satisfying from the view of experimental confirmation, it was compatible through its ontologies. For further details, refer to \cite[]{kuhn1957copernican}.
\newline
\newline
The OCA points in the direction of scientific realism, where the ontology of a theory plays a crucial role, and its connection to empiricism, instrumentalism, phenomenalism and constructivism will be investigated in a formal language in a subsequent work.


\section{Ockham and the interpretations of Quantum Mechanics}

There is a long tradition in struggles around the interpretation of Quantum Mechanics, beginning in the 1920\textsc{\char13}s, to argue with principles. One principle we will discuss here, which showed up pretty late on the stage of interpretational discussions, is Ockham\textsc{\char13}s razor. Its appearance can be taken as an indication that the discussion somehow stagnated due to a lack of new experimental evidence, obdurate positions and a new generation of physicists. 

In the following, we will present a collection of citations from papers and publications where Ockham\textsc{\char13}s razor was applied. 
We will first present a short overview of the interpretations involved, present the citations by order of their positions, summarize the views and draw our conclusions.

\subsection{Overview of interpretations used in the discussion}

\subsubsection{Copenhagen interpretation} 
The Copenhagen interpretation represents the view that the wavefunction, or the state vector, of a system gives a complete description of the physical situation of a quantum mechanical system. The evolution of the wavefunction or state vector is given by the Schr\"odinger equation. The measurement process of a physical quantity is not described by the Schr\"odinger evolution, it is defined by the projection postulate which assigns measurement outcomes to eigenvalues of specific operators. This process is often called the collapse of the wavefunction, because during a measurement process the unitary evolution of the wavefunction is replaced by this non-unitary collapse. According to Copenhagen interpretation, this collapse occurs objectively random and the physical quantities have no definite values before the measurement process. Furthermore, according to the Copenhagen interpretation, it has no meaning at all to speak about the values of a physical systems quantities prior to the measurement.  The Copenhagen interpretation is therefore a indeterministic\footnote{Indeterminism is the position that an event has no causal reason; it is the opposite of determinism, where every event has causal reason. Both positions are connected to the concepts of probability and free will.}, non-realistic\footnote{In this context, realism means that properties and quantities of physical systems have a meaning and definite value prior to their measurement. Nonrealistic interpretations would deny this assumption and claim that it makes no sense to speak about quantities of physical systems prior to measurement.} interpretation of Quantum Mechanics. 

\newpage
\subsubsection{De Broglie-Bohm interpretation}
The De Broglie-Bohm interpretation is a realistic, deterministic, nonlocal\footnote{Locality is the concept that events are independent of all events that are space-like separated. Nonlocality abolishes this restriction and, therefore, events can have an instantaneous influence on space-like separated events.}  hidden-variable interpretation and is based on the fact that the Schr\"odinger equation can be interpreted as set of one real- and one complex valued equation. The complex-valued equation is interpreted as the conservation of probability, and the real-valued equation is interpreted as a Hamilton Jacobi equation, with an extra term. This term is interpreted as a quantum potential. From the Hamilton Jacobi equation an equation of motion for the particles can be derived, which is called guidance equation. In the De-Broglie Bohm interpretation, nonlocality is an important and inherent feature of Quantum Mechanics.
 
\subsubsection{Bohmian Mechanics}
Advocates of Bohmian Mechanics claim that it is an independent theory based on the ideas of the De Broglie-Bohm interpretation, but it postulates the guidance equation for particles. In that view, it is not seen as an interpretation of Quantum Mechanics, and the guidance equation is not seen as a consequence of the Schr\"odinger equation. Bohmian Mechanics claims to be identical in all experimental predictions with standard Quantum Mechanics and is a realistic, deterministic, nonlocal hidden-variable theory. Its fundamental entities are the particles and their positions.

\subsubsection{Many Worlds interpretation}
The Many Worlds interpretation of Quantum Mechanics is a realistic deterministic interpretation that claims that the wave function is a complete description of the quantum mechanical processes. Therefor no collapse of the wavefunction hypothesis is necessary to explain the measurement process. During the measurement process a specific branch of the wavefunction emerges through the measurement interaction.

\subsubsection{QBism}
QBism or Quantum Bettabilitarianism \cite[]{fuchs2016qbist} is a continuation of the Copenhagen interpretation. It is the attempt to reconstruct Quantum Mechanics on the basis of probabilities as representation of the subjective degrees of belief of agents interacting with the world. The quantum state is not seen as a foundational entity, it is merely a different way to represent the probabilities. 
The Hilbert space is constructed out of SIC-POVMs which are symmetric, informationally complete, positive operator-valued measures. 
They generate the probabilities for the individual agents based on their subjective beliefs according to the fundamental postulate that reproduces the Born rule for the probabilities. \newline QBism is a local theory, takes a realistic position (for a detailed account on participatory realism, refer to \cite[]{fuchs2017participatory}) and provides an epistemic interpretation of the wavefunction. 

\subsubsection{Mathematical Universe Hypothesis}
The MUH is a hypothesis and not an interpretation of Quantum Mechanics, but it is connected to the Many Worlds interpretation, since the Many Worlds interpretation is contained in the MUH. The MUH claims that mathematical objects form the only external reality. It naturally explains Wigner\textsc{\char13}s Unreasonable Effectiveness of Mathematics \cite[]{wigner1960unreasonable}.

\subsection{In favor of De Broglie-Bohm Interpretation and Bohmian mechanics - against Copenhagen interpretation}

Detlef D\"urr \cite[p.116]{esfeld2012philosophie} argues in favour of Bohmian Mechanics and against the application of Ockham\textsc{\char13}s razor contra Bohmian Mechanics. By standard Quantum Mechanics, he means the Copenhagen Interpretation.

\begin{quote}
\textit{``One could say, that all additional things in Bohmian mechanics, the particles and it´s trajectories, should fall victim to Ockham\textsc{\char13}s razor. We believe this is a wrong conclusion.
We believe that ontology is nothing secondary, contrary it should be at the beginning of a theory, and that Ockham\textsc{\char13}s razor leaves behind a mutilated incomplete description of things.''}
\end{quote}

or 

\begin{quote}
\textit{``On one hand you have to accept that the description of measurements need rules that are completely different from all other laws even and cannot be derived from them. Already here fails the argument of Ockham\textsc{\char13}s razor, standard quantum mechanics is not more economic than Bohmian mechanics.''}
\end{quote}

Lucien Hardy argues in favour of hidden variables \cite[]{cushing2013bohmian}:

\begin{quote}
\textit{``It is very often argued that Ockham\textsc{\char13}s razor should be applied here to rule against introducing such hidden variable, but this rather depends on how one interprets the objective of applying Ockham\textsc{\char13}s razor and the need to provide explanation. There are perhaps three reasons why one might want to introduce hidden variables into quantum mechanics: (1) to restore determinism (2) To provide a clear ontology (3) To solve the measurement problem.''}
\end{quote}

Oliver Passon indicates in \cite{passon2004isn} that Ockham\textsc{\char13}s razor is used in the wrong direction and would actual favour Bohmian Mechanics: 

\begin{quote}
\textit{``...the deBroglie-Bohm theory supplements ordinary quantum mechanics by an equation-of-motion for the quantum-particles, but eliminates the postulates which are related to the measurement process(not to mention how uncompelling those postulates are).''}
\end{quote}

and further 

\begin{quote}
\textit{``Hence, it is questionable whether the precondition for applying Ockham\textsc{\char13}s razor is met.''}
\end{quote}

It is unclear what precondition \citeauthor{passon2004isn} here refers to. The message still is clear; to him, Ockham\textsc{\char13}s razor cannot decide in favour of Copenhagen interpretation.

\subsection{In favor of Copenhagen interpretation - against De Broglie-Bohm Interpretation and Bohmian Mechanics}

One more example of how Ockham\textsc{\char13}s philosophy is used to argue in favour or against a certain interpretation comes from Anton Zeilinger \cite[p.151,152,154]{zeilinger2003einsteins} when he speaks about Ockham\textsc{\char13}s razor:

\begin{quote}
\textit{``It is the assumption, that one should not invent things, quantities or entities without necessity.''}
\end{quote}

and

\begin{quote}
\textit{``William of Ockham was a medieval philosopher who wanted to cut away all unnecessary things in philosophy.''}
\end{quote}

and

\begin{quote}
\textit{``An important argument against the quantum potential comes again with Ockham\textsc{\char13}s razor. If we can, what is indeed the case, explain just as much as without the quantum potential, then it is redundant.''}
\end{quote}

\cite{zeilinger1996interpretation} furthermore argues in favour of Copenhagen interpretation:

\begin{quote}
\textit{``I have purposely not dealt with questions like: Is there a border between micro- and macro physics? Is a new form of logic necessary for quantum processes? Has one's awareness an active, dynamic influence on the wave function? Such or similar positions were proposed by several physicists, but in my opinion they would all fall victim to Occam's razor: Entia non sunt multiplicanda praeter necessitatem. It is the beauty of the Copenhagen interpretation that it operates with a minimal set of entities and concepts.''}
\end{quote}

\newpage

\subsection{In favor of Copenhagen Interpretation - against Many Worlds Interpretation}

Anton Zeilinger applies Ockham\textsc{\char13}s razor also against the Many World Interpretation \cite[p.151,152,154 ]{zeilinger2003einsteins}:

\begin{quote}
\textit{``The Many-worlds interpretation violates a basic assumption, which has been very successful in the past. … Why should one invent so many worlds, if there are other interpretations, which do the same without them? ''}
\end{quote}

Nicolas Gisin \cite[p.5]{gisin2013there} argues against Many Worlds, but unfortunately the reference in the paper refers not to the resource of the original critique: 

\begin{quote}
\textit{``Years ago, I once argued that the many-worlds does not seem compatible with Occam's razor principle. As answer I got the following:  ``Occam's razor should not be applied to the physical world,  but  be  applied  to  the  Schroedinger  equation;  don't  add any term to this beautiful equation''.  The linearity of the
Schroedinger equation was assumed more real than our physical universe.''}
\end{quote}

\subsection{In favor of Many Worlds Interpretation - against De Broglie-Bohm Interpretation and Bohmian Mechanics }

Hilary Greaves argues with Ockham\textsc{\char13}s razor in favour of Many Worlds Interpretation against the hidden variable approaches, which adresses the De Broglie-Bohm Interpretation and Bohmian Mechanics \cite[]{greaves2007probability}:

\begin{quote}
\textit{``The motivations for the many-worlds interpretation include (i) the fact that (if defensible) it solves the measurement problem, (ii) compatibility with the spirit of special relativity, and (iii) Ockham's razor. ... 
There is no 'ontological extravagance' in any offensive sense, since nothing is being added at the fundamental level. (Indeed, the fundamental ontology of this sort of many-worlds interpretation is actually more parsimonious than that of the rival 'hidden variables' approaches.''}
\end{quote}

\cite{sep-qm-manyworlds} argues in favour of the Many Worlds interpretation against Bohmian mechanics: 
\begin{quote}
\textit{``The MWI is also more economic than Bohmian mechanics ,which has in addition the ontology of the particle trajectories and the laws which give their evolution.''}
\end{quote}

\newpage
\subsection{In favor of De Broglie-Bohm Interpretation and Bohmian Mechanics - against Many Worlds Interpretation}

\cite{valentini2010broglie} in a reply to claims that the Many Worlds interpretation includes the idea of De Broglie-Bohm Interpretation:

\begin{quote}
\textit{``However, since the de Broglie velocity field is single-valued, trajectories $q(t)$ cannot cross. There can be no splitting or fusion of worlds. The above de
Broglie-Bohm multiverse then has the same kind of `trivial' structure that would be obtained if one reified all the possible trajectories for a classical test particle
in an external field: the parallel worlds evolve independently, side by side. Given such a theory, on the grounds of Ockham's razor alone, there would be a conclusive case for taking only one of the worlds as real.
''}
\end{quote}

\subsection{Other positions}

The following citations cannot be assigned clearly to one of the previous groups, but are relevant to our discussion. Specifically, the Mathematical Universe Hypothesis is related to the Many Worlds interpretation. In our analysis we will assign them to the same group and merge their points of view. \newline

Arthur Fine argues with a counterfactual gedankenexperiment on the development of Quantum Mechanics \cite[Arthur Fine - On the interpretation of Bohmian Mechanics]{cushing2013bohmian}:
\begin{quote}
\textit{``Given the entanglement of physics with positivistic philosophy in the first half of this century, a process initiated by Einstein and relativity, we might embellish Cushing´s counterfactual history a bit. We might conjecture that if Bohmian mechanics had indeed been the first theory of the quantum domain, subsequent epistemological discussions might well have inspired at least some participants to wield Occam´s razor to pare away everything that is not simply quantum mechanics!''}
\end{quote}

%

\cite{jannes2009some} with a critical remark on the Max Tegmarks Mathematical Universe Hypothesis (MUH):

\begin{quote}
\textit{``A classical argument against Platonism is that it unnecessarily complicates our view of reality by requiring a commitment to the existence of an immense realm of mathematical and other abstract entities, thereby violating Occam’s razor. This does not imply a denial of the effectiveness of mathematics at describing (part of) reality.''}
\end{quote}

\cite{tegmark2004parallel} with a statement concerning his Mathematical Universe Hypothesis and the application of Ockham\textsc{\char13}s razor to the Multiverse concepts. He defends the Multiverse concept against Copenhagen interpretation and along the way attacks the simplicity of the Copenhagen interpretation:

\begin{quote}
\textit{``In short, the Level III multiverse, if it exists, adds nothing new beyond Level I and Level II — just more indistinguishable copies of the same universes,
the same old storylines playing out again and again in other quantum branches. Postulating a yet unseen nonunitary effect to get rid of the Level III multiverse, with Ockham’s Razor in mind, therefore would not make Ockham any happier.''}
\end{quote}

and

\begin{quote}
\textit{``The principal arguments against parallel universes are that they are wasteful and weird, so let us consider these two objections in turn. The first argument is that multiverse theories are vulnerable to Ockham’s razor, since they postulate the existence of other worlds that we can never observe.  Why should nature be so ontologically wasteful and indulge in such opulence as to contain an infinity of different worlds? Intriguingly, this argument can be turned around to argue for a multiverse. When we feel that nature is wasteful, what precisely are we disturbed about her wasting? Certainly not “space”, since the standard flat universe model with its infinite volume draws no such objections. Certainly not “mass” or atoms either, for the same reason — once you have wasted an infinite amount of something, who cares if you waste some more? Rather, it is probably the apparent reduction in simplicity that appears disturbing, the quantity of information necessary to specify all these unseen worlds. However, as is discussed in more detail in Tegmark (1996), an entire ensemble is often much simpler than one of its members.''}
\end{quote}

\cite{fuchs2010qbism} remarks concerning a specific aspect of QBism:

\begin{quote}
\textit{``The aim of physics is to find characteristics that apply to as much of the world in its varied fullness as possible. However, those common characteristics are hardly what the world is made of - the world instead is made of this and this and this.''}
\end{quote}

This statement does not address Ockham\textsc{\char13}s razor, but it points out the position of Ockham in the debate on universals; it is a nominalists point of view. It is worthwhile to analyse if the characteristic of a position in the debate on universals can be found in other interpretations as well.  The scope would be to answer whether it would take a realist position, which means in the context of the problem of universals, a neutral position, and if this characteristic can be used as a differentiation between interpretations. This will be done in a subsequent paper.

\newpage

\subsection{Analysis of applications of Ockhams razor}

The presented meta-analysis cannot claim to be complete,as it did not cover all journals or even a nameable subset of all relevant books in this field; nor can it be used to draw quantitative conclusions, though it is a representative overview of the status quo in the literature, reflects a remarkable situation and allows us to draw conclusions. \newline

\begin{figure}[h] 

\begin{center}
\includegraphics[width=0.6\textwidth]{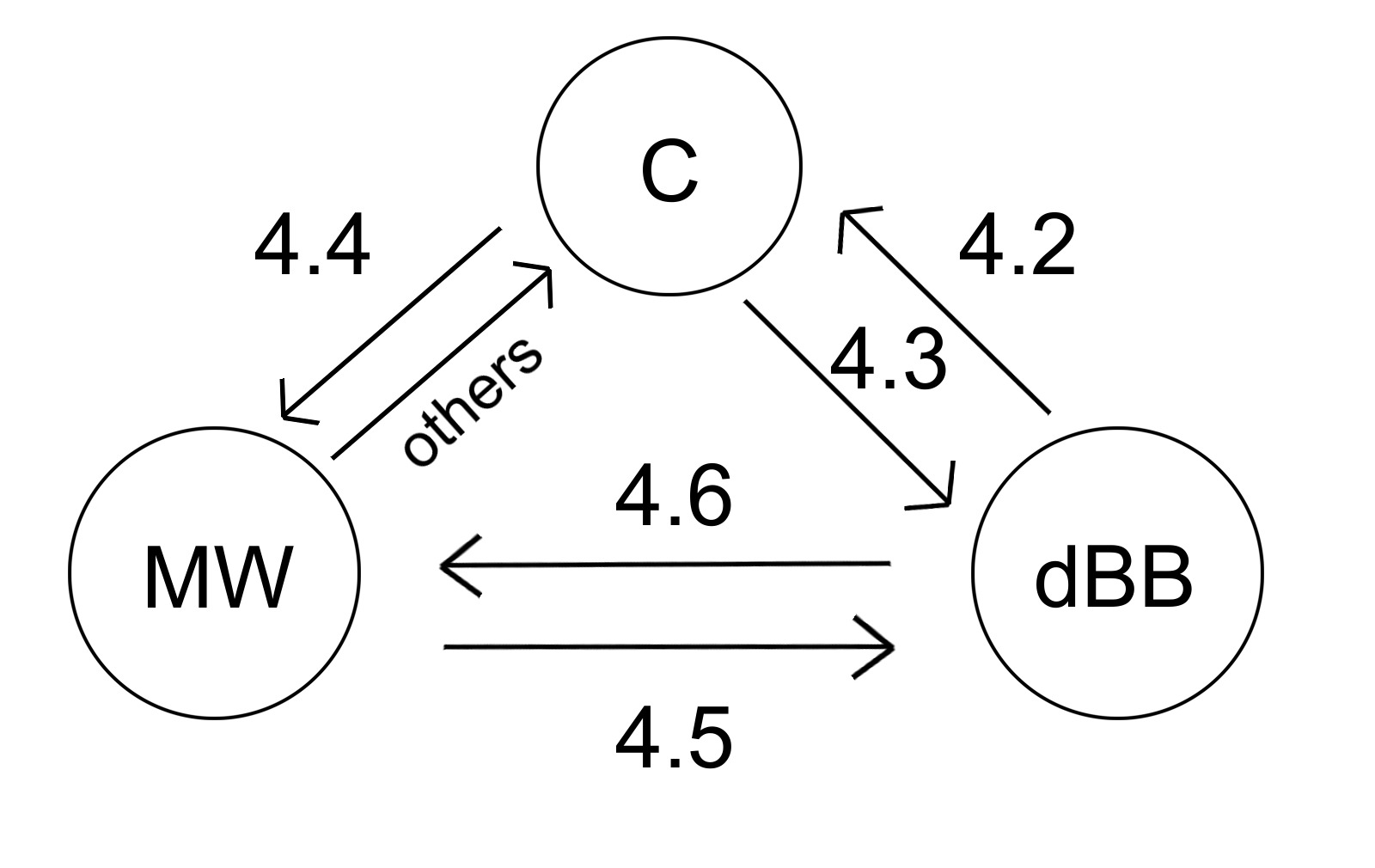}
\caption{Contradictions in application of Ockham\textsc{\char13}s razor} 
\label{Contradictions}   
\end{center}

\end{figure}

In the above meta-analysis, Ockham\textsc{\char13}s razor is used in nearly all constellations by proponents of all positions. Positivist against realist point of view, realist against positivist point of view, deterministic interpretation against indeterministic interpretation, orthogonal interpretations arguing against each other. Figure \ref{Contradictions} shows the positions and their respective contradictions. It turns out that they are all in contradiction to each other. 

How is it possible that a principle that is meaningful to several advocates of different viewpoints in the interpretational debate allows such contradictory conclusions?

One of the problems of the application of Ockham\textsc{\char13}s razor in general is that the principle itself argues in favour of parsimony over plentifulness, in favour of simplicity. And there it becomes difficult in our application to the interpretations of Quantum Mechanics. As we have seen, the definition of simplicity is not a standard one; even more in the above-cited viewpoints, it is rarely argued what is meant by this simplicity, as if it was obvious. A resolution for this could be to define a complexity measure of theories and then let the more simple theory according to this measure (or in our case of interpretations of Quantum Mechanics, the more parsimonious interpretation) pass. It seems that the problem then only shifts to the definition of this measure of complexity, which then will become the heavily disputed topic. One could think that this leads into an endless self-referring circle with simplicity measures that refer to simplicity. On the other hand, a measure for the complexity of an interpretation is an interesting open question that should not be unconsidered. 

\subsection{Ockham´s razor and economy of thought}

A different approach to this topic would be if we revisit the original idea of Ockham\textsc{\char13}s razor as something of common sense, as an idea of economy of thought. This idea can be found in social psychology as well. 

\cite{wegener1998naive} gives a review of the social psychology model of the naive scientist (\cite{Heider1958psychology}). The idea is that humans perceive their environment like naive scientists and make attributions to phenomena. \newline

The naive scientist model got under attack over the years and has been replaced by the cognitive miser model, which argues that humans make indeed make use of several heuristics on their judging on certain facts, but the focus lies on the efficiency of the decision process \cite[]{fiske2013social}:

\begin{quote}
\textit{``Hence the third general view of the thinker is the cognitive miser model. The idea is that people are limited in their capacity to process information, so they take shortcuts whenever they can. People adopt strategies that simplify complex problems; the strategies may not be correct or produce correct answers, but they emphasize efficiency. The capacity-limited thinker searches for rapid, adequate solutions rather than for slow, accurate solutions.''}
\end{quote}

\cite{tversky1973availability} remark on the methods of application of the different heuristics: 

\begin{quote}
\textit{``
Little is known, however about the frequency of classes or the likelihood of events. We propose that when faced with the difficult task if judging probability or frequency, people employ a limited number of heuristics which reduce these judgements to simpler ones.''}
\end{quote}

and 

\begin{quote}
\textit{``We suggest that in evaluating the probability of complex events only the simplest and most available scenarios are likely to be considered.''}
\end{quote}

We see a deep connection between the cognitive miser model in social psychology and Ockham\textsc{\char13}s razor in the way Ockham understood it, as a concept of common sense, as an economy of thought. 
\newline
\newline
\newpage

We conclude that the remarkable situation of contradictory statements concerning the simplicity of interpretations is due to the inapplicability of Ockham\textsc{\char13}s razor in the interpretations of Quantum Mechanics. This inapplicability reflects in the contradictions between the different viewpoints in the interpretations of Quantum Mechanics, which shows that there is no preferred direction in how Ockham\textsc{\char13}s razor is used, and it seems that it is up to subjectivity as to how it is applied. It therefore seems arbitrary how Ockham\textsc{\char13}s razor is applied here at all, which allows doubts for the value Ockham\textsc{\char13}s razor can have in the discussion on the interpretations of Quantum Mechanics. \newline 
\newline
As long as there is no quantitative system that allows to account for simplicity in that regime, we are forced to rely on subjective judgements if Ockham\textsc{\char13}s razor is used reasonably in that specific case. We think that this is not an appropriate scientific method, but merely a matter of taste. As scientists, we should be very careful and alert when applying Ockham\textsc{\char13}s razor, especially if we adopt it as a first order principle, since the analysis of its historical origin showed that Ockham\textsc{\char13}s razor is just a concept of economy of thought and not a concept of simplicity of nature, which is also supported by social psychology.




\newpage
\section{A new diversity - conclusion}

We have come a long way from the medieval philosophy and its peculiarities to the disputes on identity, which are still an open question in the diverse field of the interpretations of Quantum Mechanics. We questioned the applicability of Ockham\textsc{\char13}s razor to the interpretations of Quantum Mechanics and concluded that no clear scheme can be seen as to how Ockham\textsc{\char13}s razor should  contribute in a positive way to the ongoing discussions. This is certainly owed to the notion of simplicity, which is merely a subjective quality and seems to be a relative concept. \newline


Alluding to the Great Cosmological Controversy again, we claim that there is an analogy\footnote{Though there are clearly different situations concerning the mathematical framework.}to the current situation in the interpretations of Quantum Mechanics. In the case of the cosmological models, we encounter two orthogonal world-views, but overlapping in the empirical predictions of the models \footnote{With respect to the experimental limitations of that time.} that rival for being accepted as the ``true" theory. As argued before in the discussion on the Ontological Coherence Argument, each position claimed simplicity as an argument for its world-view, underpinning it with significant arguments. \newline


It turned out that the heliocentric model delivered a world view that had striking advantages, but when relativity arose at the begin of the 20th Century the Great Cosmological Controversy was seen under a new perspective. The remarkable situation that it was not a trivial task to decide between these two models turned out to have a reason, in the fact that both models are applicable due to the relativity principle, and the question for truth if the sun or the earth rested turned out to be meaningless. In Quantum Mechanics, we are in the exceptional situation that we face several interpretations that are in full agreement with the experimental predictions but are completely orthogonal in their ontological statements and implications, even with respect to diverse aspects of the world views.\footnote{Such aspects of interpretations which show orthogonal behavior are Determinism-Indeterminism, Locality-Nonlocality, Nominalism - Platonism, Objectivity - Subjectivity, Uniqueness - Ambiguity} We conclude out of this  situation in the field of the interpretations of Quantum Mechanics that we have an analogy to the Great Cosmological Controversy. 

\newpage
 Is this situation not an indication that should encourage us to think that a dogmatic dispute with one winner might not be the solution of the problem? One can believe that the current situation in Quantum Mechanics can be understood that way and that this situation can be interpreted as a perspective on the new physics that may lie ahead, indicating that our well-established concepts of quantum-objects, spacetime, determinism, probability and information will be challenged. Even more, it seems that our philosophy of science will be challenged. Quantum logic showed that even the propositions of logic can be involved in the ambiguous story of Quantum Mechanics \cite[]{mittelstaedt1978quantum}. To underpin the importance of this ambiguousness, we call this quality of Quantum Mechanics ``Quambiguity"   \newline

\begin{figure}[h] 

\begin{center}
\includegraphics[width=0.6\textwidth]{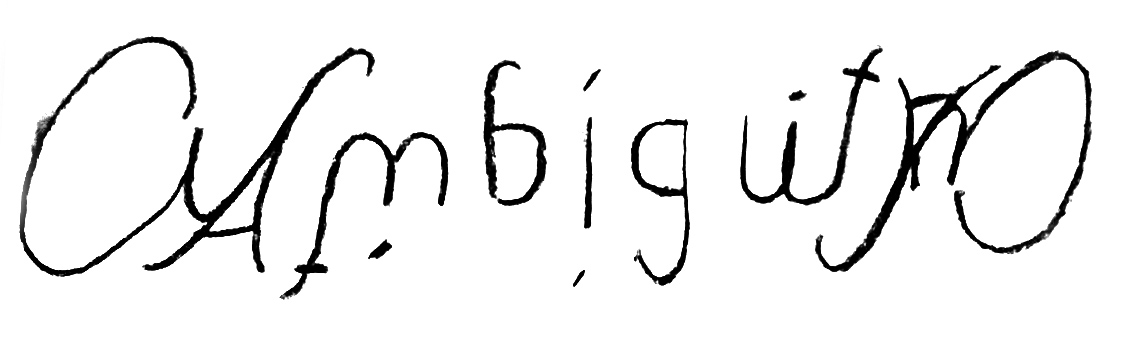}
\caption{Quambiguity}
   
\end{center}

\end{figure}

In any case, we want to conclude that Ockham\textsc{\char13}s razor cannot solve this mystery in a swift way. It is a concept of parsimony of thought, but was not meant to give a guideline that nature has to be simple in principle. It originates in the economy of thought that is an implication of the limited resources of our cognitive skills, which reflects in well-known heuristics in social psychology. For Ockham himself, the principle was a heuristic of common sense that helped him in his line of argument against Platonism, against a realist position in the debate on the universals. He used the razor to get rid of a not necessarily needed multiplicity of entities.  \newline

In context with Quantum Mechanics, Ockham\textsc{\char13}s razor should encourage us to reconsider the principles and mechanism of our thinking and reasoning, our judging about certain theories, world-views and mindsets, and our prejudices.  

\section{Acknowledgments}

I would like to thank the head of the Quantum Particle Workgroup Beatrix Hiesmayr for her support, and Chris Fuchs, Stefanie Lietze, Agnes Rettelbach, Johnjoe McFadden and several other colleagues for fruitful discussions. I would like to thank Josef Reiter for supporting the translation of the Latin citations. For the realisation of the graphical elements and illustrations, I would like to thank Isabella W\"ober.

\newpage

\bibliography{G:/Physik/Bibfiles/Quantenmechanik}

\appendix
\newpage

\section{Appendix - Arguing with the scholastic method}

Using Ockham\textsc{\char13}s discussion on universals from his Opera Theologica II \cite[OT II, Liber I, D.2 Q.4, page 100]{william2012william} the structure of the scholastic tradition can be seen quite well in the following example. \\ \\
We follow the translation from \cite[]{spade1994five}. Ockham there poses the question: 

\begin{quote}
\textit{``Question 4: Is a universal a thing outside the soul, in individuals, really distinct from them, and not multiplied in individuals?''}
\end{quote}

Then he argues in favour of the position that affirms the question:

\begin{quote}
\textit{``(1) As for the identity and distinction of God from creature, it must be asked whether there is something univocal common to God and creature and essentially predicable of both. But because this question, along with much of what has already been said and is about to be said in the following questions, depends on a knowledge of univocal and universal nature, there fore in order to clarify what has been and will be said I will first ask some questions about universal and univocal nature. \\
(2) On this, I first ask whether what is immediately and proximately denominated by a universal and univocal intention [K Ch. 23] is truly some thing outside the soul, intrinsic and essential to what it is common and univocal to, and really distinct from them. 
(3) Yes it is: \\
(4) First, it is truly a thing, essential and intrinsic to what it is common to. \\
...''}
\end{quote}

He then presents the view of his opposer, who he calls  \textit{``Doctori Subtili''}\cite[OT II, Liber I, D.2 Q.4, page 100]{william2012william}. Some researchers assume that he refers to Duns Scotus, and some would argue that the view he presents would better represent Walter Burleys ideas \cite[page 115]{spade1994five}:  

\begin{quote}
\textit{`` Walter Burleys view - Opinio Doctori Subtili Imposita \\ \\
(8) On this question there is one theory that says every univocal universal is a certain thing existing outside the soul, really in each singular and belong ing to the essence of each singular, really distinct from each singular and from any other universal, in such a way that the universal man is truly one thing outside the soul, existing really in each man, and is really distinguished from each man and from the universal animal and from the universal sub stance.\\
...''}
\end{quote}

Following an extensive argumentation in favour on this view a not less extensive list of arguments against it is presented:

\begin{quote}
\textit{``Against this theory - Contra Opinonem Doctori Subtili Impositam \\ \\
(34) This view is absolutely false and absurd. So I argue against it. \\
(35) First: No thing one in number, not varied or multiplied, is in several sensible supposita or singulars, or for that matter in any created individuals, at one and the same time. But a thing such as this theory postulates, if it were granted, would be one in number. Therefore, it would not be in several singulars and belong to their essence. \\
...''}
\end{quote}

Now Ockham presents his own views

\begin{quote}
\textit{``Ockham\textsc{\char13}s own theory - Opinio Auctoris \\ \\(95) Therefore, in reply to the question [(2)], I say otherwise: No thing really distinct from singular things and intrinsic to them is universal or common to them. For such a thing is not to be posited except (a) to preserve the one's essential predication of the other, or (b) to preserve our knowledge of things and (c) the definitions of things. Aristotle, [Metaphysics XIII.4, 1078,27–34], suggests these reasons for Plato's theory.\\
...''}
\end{quote}

followed by replies to the before presented theories:

\begin{quote}
\textit{``Replies to the arguments in favor of Burley's theory - Ad Argumenta Opinionis Doctori Subtili Impositae \\ \\
(111) To the other theory’s first argument \\
...''}
\end{quote}

and

\begin{quote}
\textit{``Reply to the preliminary arguments - Responsio Ad Argumenta Principalia \\ \\
(184) To the first main argument [(4)], I reply: ‘The universal man and a particular man are essentially one is literally false.\\
...''}
\end{quote}

The scholastic method is a completely different way of arguing than what we are used to nowadays. To modern eyes, it appears in some sense excessive and inefficient, though it prepared the ground for modern scientific reasoning, which is built on hypotheses, induction, deduction, and discussion. 

For more details concerning the scholastic method, refer to \cite[]{kretzmann1982cambridge}. There, a summary and detailed presentation of the scholastic method and its historical embedding is given.

\section{Appendix - Modern views on the Contradiction principle, the notion of identity and indistinguishability}

Ockham\textsc{\char13}s Contradiction principle has some predecessors and modern representations, and pervaded the history of philosophy in different representations and names. It would exceed the scope of this paper to give an overview that could claim to be complete. Therefore, a certain selection is presented that seems particular relevant to the author in context of this discussion to demonstrate the application of Ockham\textsc{\char13}s razor. \newline

\citeauthor{whitehead1912principia} are prominent for their attempt to put mathematics on a logical basis. In this canon, the law of Non-Contradiction also found its place

\cite[p.117 *3,24]{whitehead1912principia}:

\begin{quote}
\textit{$\vdash \lnot(p \wedge \lnot p )$}
\end{quote}
Two contradictory propositions cannot both be true. In the context of propositions about the properties of objects, difference can be posed through contradiction. This argument and application of Contradiction is found later in  \cite{wittgenstein1994tractatus} which refers to this principle several times. See \cite[2.0233 and 2.02331]{wittgenstein1994tractatus}:

\begin{quote}
\textit{``If two objects have the same logical form, the only distinction between them, apart from their external properties, is that they are different.'' }
\end{quote}

\begin{quote}
\textit{``Either a thing has properties that nothing else has, in which case we can immediately use a description to distinguish it from the others and refer to it; or, on the other hand, there are several things that have the whole set of their properties in common, in which case it is quite impossible to indicate one of them.
For if there is nothing to distinguish a thing, I cannot distinguish it, since otherwise it would be distinguished after all.''}
\end{quote}

or at a different place \cite[5.53, 5.5301, 5.5302, 5.5303]{wittgenstein1994tractatus}:

\begin{quote}
\textit{``Identity of the object I express by identity of the sign and not by means of a sign of identity. Difference of the objects by difference of the signs.'' }
\end{quote}

\begin{quote}
\textit{``It is self-evident that identity is not a relation between objects. This becomes very clear if one considers, for example, the proposition $(x):fx.\subset.x=a$. What this proposition says is simply that only $a$ satisfies the function f, and not that only things that have a certain relation to $a$ satisfy the function. Of course, it might then be said that only $a$ did have this relation to $a$; but in order to express that, we should need the identity-sign itself.'' }
\end{quote}

\begin{quote}
\textit{``Russell's definition of '=' is inadequate, because according to it we cannot say that two objects have all their properties in common. (Even if this proposition is never correct, it still has sense.)'' }
\end{quote}

\begin{quote}
\textit{``Roughly speaking, to say of two things that they are identical is nonsense, and to say of one thing that it is identical with itself is to say nothing at all.'' }
\end{quote}

The laws of contradiction are always in connection with the question of identity of objects, beginning with the pre-Socratics and their philosophy of substance and matter. Aristotle based this and, of course, his own philosophical ideas on logical foundations that still pervade our thinking. In classical physics, the identity of physical objects seemed clearly defined, but the foundational questions have there have been already covered by the operational success Newtonian Mechanics offered. When Quantum Mechanics arose, those philosophical questions became significant again in physics because of their sudden experimental accessibility and support by measurements on the microscopical scale. 

The individuality and nature of quantum objects became a striking question of quantum philosophy, starting at the statistical mechanics involving Quantum Mechanics. By neglecting the identity of permuted particles, the experimentally confirmed laws of statistical mechanics have been found. It is an interesting observation that physics and philosophy develop in a kind of entangled way and stimulate each other. 
\newline

Quantum mechanics kept the topic of identity challenging up to now, as can be seen by ongoing discussions on the definition and interpretation of identity, individuality and indistinguishability \cite[]{krause2014separability}; \cite[]{daCosta2014quantum};\cite[]{dieks2014logic}. For an account of individuality in the Bohm interpretation, refer to \cite[]{pylkkanen2014bohm}.\newline

\end{document}